\documentstyle[prl,aps,multicol,epsf]{revtex} \tighten 
\begin{document} \draft
\title{Electrophoretic separation of proteins via
complexation with a polyelectrolyte}
\author{E. M. Baskin $^1$ , B. I. Shklovskii $^2$
 and G. V. Zilberstein $^3$ }
\address{ $^1$ Department of Physics and Solid State Institute, 
Technion - IIT, Haifa 32000, Israel\\ $^2$ Theoretical Physics
Institute, University of Minnesota,
116 Church St. Southeast, Minneapolis,
Minnesota 55455\\
$^3$ Protein Forest Inc., 4 Pekeris st., Rehovot 76701, Israel}

\maketitle

\begin{abstract}
We suggest to augment standard isoelectronic focusing
for separation of proteins in a gradient of pH
by a similar focusing in the presence of
a strongly charged polyelectrolyte (PE).
Proteins which have the same 
isoelectric point but different "hidden" 
charge of both signs in pI point 
make complexes with PE, which focus in 
different pH. This is a result of charge
inversion of such proteins by 
adsorbed PE molecules, which is 
sensitive to the hidden charge.
Hence the hidden charge is a new separation 
parameter.

\end{abstract}

\begin{multicols}{2}

Separation of proteins in
a gradient of pH, or isoelectric focusing 
is the basis of proteomics\cite{R1,R2}. 
It uses the fact that 
proteins have  
both basic and acidic groups.
Let us assume that on 
the left side of electrophoretic cell
pH is so small that all basic groups are positively charged
and all acidic ones are neutral,
 so that the protein is net positive and 
moves in electric field to the right 
where pH increases. While protein moves to the right 
more and more basic groups are neutralized and some 
acidic one dissociate and become negative.
As a result for each protein there is the so called 
isoelectric point pH = pI,
where its charge changes sign. Beyond this point 
a protein becomes negative and returns back. 
Thus, each protein accumulates near its pI. 
It is not, however, totally satisfactory because there are many 
different proteins with the same pI. 
One still needs another parameter to separate them. 

A standard method to do this consists in 
addition of a detergent SDS,
which has a long hydrophobic 
tail and a negatively charged head\cite{SDS}. 
Molecules of SDS denature a protein 
globule by attaching their tails to hydrophobic 
parts of protein and cover it by charged heads. 
Resulting rod-like negatively charged complexes are separated 
by their mobility, which in turn depends on the length 
of the protein and the number of its hydrophobic groups. 
This is the second coordinate 
of two-dimensional electrophoresis of proteins (pH is the first). 

In many cases even this two-dimensional analysis 
does not provide necessary resolution.
Furthermore, native structure of proteins is 
irreversibly destroyed by SDS and the protein can not 
be used for a farther analysis.
Therefore, if possible, another method should be used together with 
isoelectric focusing.

In this paper, we suggest a different idea 
for separation of proteins with the same pI.
In addition to the total charge
of protein this method is sensitive mostly to the 
absolute value, $q$, of the positive and negative
"hidden" charges at its pI. 
Our goal is to separate proteins 
with the same pI, but different $q$.
For this purpose, we suggest to complex protein globules with a
strongly charged,  
short polyelectrolyte (PE). 
For different proteins with the same pI
protein-PE complexes 
have a different number of PE molecules which depends on $q$.
Correspondingly, isoelectric points of these complexes 
(values of pH where they are neutral)
differ from a protein to protein. Therefore, 
complexes of different proteins can
be separated by isoelectric focusing in 
a gradient of pH. 
Using short PE is important because a long one could 
bind several different proteins together. 

Let us for a given protein 
plot on Fig. 1 
the isoelectric points of the protein (thin line)
and protein-PE complex (thick line)
as a function of the concentration 
of PE, $N$. 
Adding to this plot the signs of the bare protein 
and the protein-PE complex
we obtain the "sign" phase diagram in the plane (N, pH)
\begin{figure}[h]
\epsfxsize=8cm \centerline{\epsfbox{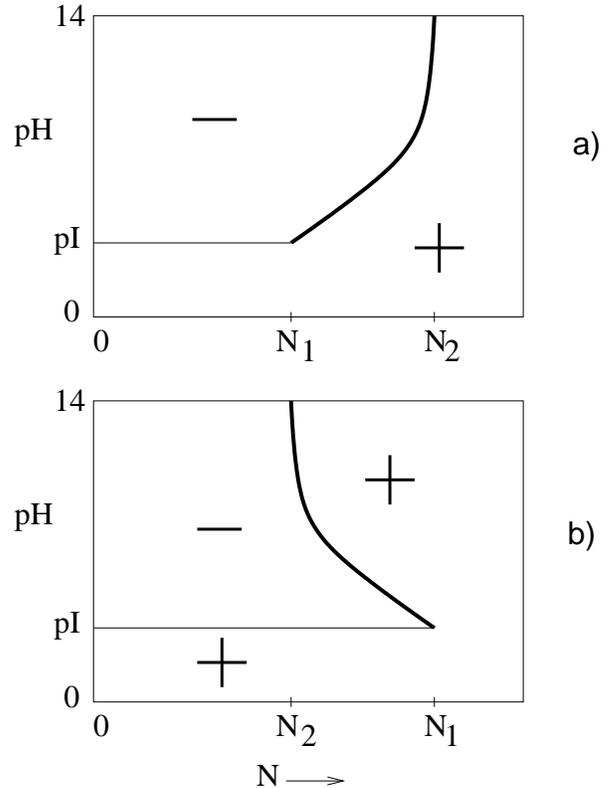}}
\caption{The sign phase diagram in 
the plane (N, pH) for the scenarios A (a) and B (b).
The horizontal thin line corresponds to pH = pI, 
the curve pH = pI$(q, N)$ is shown by the thick line.}
\label{fig:model}
\end{figure}
In the case of small enough $N$,
when there are no complexes at all,
plus and minus are separated by the horizontal thin line pH($N)$ = pI.
We show below that 
there are two different scenarios
for deviation of the thick line 
from the horizontal line, when complexes
appear. They lead to two different "sign"
phase diagrams. According to 
the scenario A (Fig. 1a), 
deviations from horizontal straight line happens
when concentration 
of PE, $N$, reaches the critical concentration $N_1$
at which the neutral protein adsorbs the first molecule of PE.
At $N > N_1$ this complex is neutral at pH = pI$(q, N) > pI$. 
Thus, at $N = N_1$ the horizontal line pH = pI, crosses over to 
pH = pI$(q, N)$ curve, which deviates in the direction of larger pH.   
At some larger $N = N_2$ the point pI$(q, N)$ reaches pH =14. 
The complex is positive at pH $ <$ pI$(q, N)$ and negative 
at pH $>$ pI$(q, N)$ (Fig. 1a).

The scenario B is different from 
A, because 
in this case $N_1 > N_2$ (see Fig. 1b). 
In this scenario again the neutral protein at pH = pI does not 
adsorb a PE molecule till $N = N_1$.
However, at pH = 14 where 
protein is strongly negatively 
charged the charge of the complex can become positive 
already at smaller concentrations $ N_2 < N < N_1 $.
As a result at a fixed $N$ in the range of concentrations
$ N_2 < N < N_1 $, protein changes sign twice with growing pH (Fig. 1b). 
First time this happens in the standard pI point of the bare protein. 
At pH $>$ pI  protein becomes negative, so that at some point it starts
adsorbing PE molecules. If pH increases further at some point  
pI$(q, N)$ protein charge becomes neutralized by 
adsorbed PE molecules. At even larger pH the 
bare charge of protein 
is so large that it becomes overcharged by PE.

In both scenarios the new isoelectric point pI$(q, N)$
depends both on $q$ and $N$.
In other words, it is different for
protein-PE complexes of different proteins with the same pI.
This is the basis for proposed method of 
separation of proteins with the same pI.
Although we literally defined $q$ as a hidden charge
one can also think that this notation in formula pI$(q, N)$
includes other hidden parameters which 
discriminate between different proteins with the same pI.

In a simple-minded experiment one can cut out 
a stripe of the gel, where proteins with a given 
pI and different $q$ are focused
and put into another device with 
orthogonal to the stripe pH gradient
and with the concentration $N$ of a PE. 
In this paper, we have in mind 
this simple set up although
it is possible that in future
one can create a gradient 
of $N$ orthogonal to the gradient of 
pH and visualize Fig. 1. 
 
Before explanation
of the origin of the scenarios A and B 
let us emphasize the common nontrivial 
feature of these scenarios.
Both of them result from the
phenomenon of charge inversion of a protein by PE.
Let us consider it on the example of 
large pH close to 14, when
all basic groups of the bare 
protein are neutralized and all acidic ones are 
ionized so that the bare 
protein has a large negative charge. 
Nevertheless, if the concentration of added PE, $N$, is large enough,
the total charge of PE molecules sticking to the protein 
can be even larger than bare charge so that the 
net charge of the protein-PE complex 
is positive. In other words,
a strongly negatively charged bare protein 
already neutralized 
by adsorbed PE molecules 
continues to attract new PE molecules.
This counterintuitive phenomenon is called charge 
inversion. It becomes possible
because PE molecules repel each other and
form a correlated liquid at the surface of the protein.
A new PE molecule approaching the neutral protein-PE complex repels 
already adsorbed molecules on the protein surface
and creates a correlation hole in this liquid, which plays the role 
of the electrostatic image of PE. As a result the new PE molecule
is attracted to its image\cite{Shklov,Netz,Nguyen,RMP}.
When $N$ is large enough this attraction becomes 
more important than the loss of entropy due to adsorption.
Thus, at a given pH$>$pI the charge of the complex changes its sign 
as a function of growing $N$. This phenomenon is obvious
on both phase diagrams of Fig. 1, when at a given pH 
we cross the thick line while increasing $N$.

In a more quantitative language, one can say that 
image forces 
lead to the negative correlation chemical potential of a 
new PE molecule on the surface of already neutralized protein, 
$\mu_{s}(q, 14)$. If  $\mu_{s}(q, 14) <  \mu_{b}(N)$,
where $\mu_{b}(N)$ is the (negative) chemical potential of a 
PE molecule
in the bulk of solution, the negative bare protein
becomes overcharged by adsorbed positive PE molecules
so that the whole complex becomes positive. This can happen 
at pH$ < $14 as well.

Charge inversion by a PE is used in the gene delivery in
order to invert the negative charge of DNA
by a positive PE. 
This facilitates DNA penetration through a 
negative cell membrane\cite{KK}.
The change of the sign of DNA with increasing $N$
was recorded by the sign change of electrophoretic mobility\cite{KK}.
(See also the recent review paper on physics of charge inversion\cite{RMP}
and references therein).

Now we can explain the origin of the
difference between the scenarios A and B.
Let us discuss what happens when pH increases 
from  pI to 14 for a protein with a given $q$. 
Consider the surface chemical potential of a PE
molecule adsorbed to a {\it neutral} 
protein-PE complex, $\mu_{s}(q,$ pH),
or, in other words, free energy of 
adsorption of an additional PE
molecule by the neutralized complex. 
We already emphasized that at pH 14 
the chemical potential 
$\mu_{s}(q,$ 14) is negative. 
Actually it is natural to assume that 
$\mu_{s}(q,$ pI) is negative, too. Indeed, 
at pH = pI there are $q$ positive and $q$ 
negative charges more or less
randomly distributed at the surface 
of the protein globule.
A PE molecule can be adsorbed at the surface
due to spatial fluctuations of this charge. 
Positive monomers of a
PE molecule can approach preferably negative charges of the
surface avoiding positive ones. 
Of course, both the intrinsic rigidity of PE 
and the electrostatic rigidity
due to repulsion of PE charges 
do not let the PE molecule to avoid all positive 
surface charges. The PE molecule chooses
a spacial scale of fluctuations of the surface charge 
which it can adjust to without too much of 
loss of the elastic energy. As a result of this
optimization the PE molecule can bind to the neutral protein.

Thus we see that both $\mu_{s}(q,$ pI) and $\mu_{s}(q, 14)$
are negative. Now we have to consider two cases:
A) $\mu_{s}(q,$ pI) $< \mu_{s}(q, 14)$ and 
B) $\mu_{s}(q,$ pI) $> \mu_{s}(q, 14)$.
They generate the two above mentioned scenarios A and B.

In the scenario A, when $N$ and $\mu_{b}(N)$ grow,
at the concentration $N = N_1$, where 
$\mu_{b}(N_1) = \mu_{s}(q,$ pI), the first PE 
molecule is adsorbed by the protein.
Therefore, at $N > N_1$ we deal with the isoelectric 
point of the protein-PE complex, pI$(q, N)$. 
Positive charge of adsorbed PE molecules
shifts pI$(q, N)$ to pH$ > $pI in order to add the 
compensating negative
charge to the protein itself. 
The number of adsorbed PE  
molecules grows with $N$ and pushes pI$(q, N)$ 
to larger and larger values. At $N = N_2$,  
when $\mu_{b}(N_2) = \mu_{s}(q, 14)$ isoelectric point
pI$(q, N)$ reaches its upper limit.

Let us switch to the opposite case 
$\mu_{s}(q,$ pI) $> \mu_{s}(q, 14)$, which
corresponds to the scenario B.
Consider what happens in this case with the charge 
of the protein complex when $N$ grows (see Fig. 1). 
There are again two characteristic concentrations 
$N_1$ and $N_2$, defined by the same equations 
$\mu_{b}(N_{1}) = \mu_{s}(q, $ pI) and 
$\mu_{b}(N_{2}) = \mu_{s}(q, 14)$.
In the scenario B, $N_1 > N_2$, and
therefore, there are three different ranges 
of $N$: $N < N_2$,  $N_2 < N < N_1$ and $N > N_1$.

1. $N < N_2$. If $N$ is so small that 
$\mu_{b}(N)  < \mu_{s}(q, 14)$,
then there is no charge inversion 
of the protein by PE and, therefore PE does not 
affect the focusing.
The protein is positive at 
pH $ < $ pI and it is negative at pH $ > $ pI. 

2. $N_2 < N < N_1$. In this range 
$ \mu_{s}(q, 14) < \mu_{b}(N) < \mu_{s}(q, 
$ pI$) $. When pH exceeds
pI by some finite value, the negative
protein starts to adsorb PE molecules 
but the protein-PE complex still remains 
negative. At pH = pI$(q, N)$, 
where $\mu_{s}(q, $ pH$) = \mu_{b}(N)$ the 
charge of protein-PE complex  goes through
zero and becomes positive (charge 
inversion). At pH $> $ pI$(q, N)$ 
the protein charge stays positive. 

3. $N > N_1$. For pH $>$ pI in this case, 
$\mu_{b}(N) > \mu_{s}(q, $ pH) 
so that protein adsorbs PE and has positive charge.
For pH $<$ pI the protein is always positive, too. 
There are no isoelectric points and no focusing.

We see that there is only a limited range of $N_2 < N < N_1$,
where we get the new $q$-dependent isoelectric point pI$(q, N)$, so that 
the suggested separation method can work. 
Note that for each 
direction of electric field
one of the two isoelectric points is stable, while another is unstable.
In the former case on the both sides of the point proteins
move to the isoelectric point, in the latter one they move away from it.
The standard isoelectric focusing can 
reveal only one of the two focus 
points. For example,
if positive proteins with different $q$ start on 
the left side of the device 
and drift to the right, i. e. in the direction of increasing
pH (the electric field 
is directed to the right), all proteins focus in pI. If we 
reverse the electric field 
so that the positive protein-PE complexes drift from the right 
side of the device to the left,
different proteins with the same pI focus in
the new $q$-dependent isoelectric points, pI$(q, N)$. 
Thus we can use this dependence on $q$
to separate proteins with different $q$, but with the same pI.

There is however a way to see both isoelectric points 
for a fixed direction of electric field. Let us assume that
in the absence of electric field  proteins are 
uniformly distributed in space with pH gradient.
When we apply electric field, 
they concentrate in the stable point and 
escape from the unstable isoelectric point.
As a result we should observe the maximum 
of the protein concentration in the former point and 
the minimum in the latter one.

Until now our theory was strictly phenomenological.
Actual calculation 
of chemical potentials of $\mu_{s}(q,$ pI) and $\mu_{s}(q, 14)$,
which helps to choose between the scenarios A or B,
requires a detail theory, which 
takes into account the distribution of protein aminoacids,
the shape of the protein globule,
flexibility of PE and the protein, 
and their geometrical dimensions. 

Below we give some microscopic estimates
of $\mu_{s}(q,$ pI) and $\mu_{s}(q, 14)$
in a toy model,
which is not reliable enough to choose
between scenarios A and B for real proteins and PE (although
it seems that the scenario B is more likely).
These estimates, however, help 
to understand physics of $\mu_{s}(q,$ pI) and $\mu_{s}(q, 14)$  
and to explain the origin of their dependence on $q$,
which leads to possibility of protein separation. 

Let us consider a toy model
of a protein of approximately 200 aminoacids 
as a rigid sphere with radius $R  = 2$~nm and
$q = 20$ positive and negative charges randomly
distributed at the surface
of the sphere. 
We assume that PE is strongly charged
with the linear density of the order of 
$\eta = e/l_B$, where $e$ is the proton charge,
$l_B = e^{2}/\varepsilon k_BT \simeq 0.7$~nm 
is the Bjerrum length, $\varepsilon \simeq 80$ 
is the dielectric constant of water. 
Assume that PE has $Z=4$ charges, so that its
length $(Z-1)l_B$ is comparable to $R$. 

We concentrate on the estimates of $\mu_{s}(q, $ pI) and 
$\mu_{s}(q, 14)$ and their comparison.
Let us first estimate $\mu_{s}(q, 14)$.
At pH 14 the charge of the globule $-q$ is large, so that
protein attracts a number of PE molecules. 
A protein with $q = 20$ is neutralized by five
PE molecules with the charge $Z = 4$. The distance between 
nearest adsorbed PE molecules is of the order of $1.5R$.
Let us assume that the screening 
radius of monovalent salt, $r_s = 1~$nm, i.e. 
it is three times smaller than the distance 
between PE molecules. Then, in the 
first approximation,
one can neglect  energy of screened repulsion
between adsorbed PE molecules 
and calculate $\mu_{s}(q, 14)$
as the energy of attraction of
PE molecule to the surface of the sphere
\begin{equation}
\mu_{s}(q, 14) = - \frac{qZe^{2}r_{s}}{\varepsilon R^{2}}
 = -\frac{qZ r_{s}l_B}{R^{2}} k_{B}T.
 \label{chempot}
\end{equation}
For $q = 20$ and  $Z=4$, Eq. (\ref{chempot})
gives $\mu_{s}(q, 14) \simeq -14k_{B}T$.
This estimate disregards the fact that 
dielectric constant of the 
globule is actually much smaller
than 80. This leads to the positive 
image charge of PE inside the protein sphere under each PE.
Repulsion from the image lifts all adsorbed 
PE molecules above the protein
surface and diminishes attraction between 
PE and protein\cite{Nguyen}. The distance of PE from the protein surface 
can be calculated balancing electric field of the protein 
$qe/(R+d)^{2}$ and of the image $\eta/4d$. This gives $d \simeq 0.4~$nm.
In Eq. (\ref{chempot}) for $\mu_{s}(q, 14)$ one should now replace
$r_s$ by $r_s - d$ and also add repulsion 
energy of PE and its image 
which can be estimated as $(1/8)\eta Ze \ln(r_s/d)$.
Together these two changes lead to $\mu_{s}(q, 14) \simeq -8k_{B}T$.  
At the same time the neglected above repulsion between adsorbed PE molecules
adds another positive term. It can be calculated if we sum all 
exponentially screened repulsion energies 
of nearest neighbor PE molecules 
and then take derivative with respect
of number of PE molecules. In this procedure only
nearest neighbors should be taken into account
because of the exponential decay of the screened potential.
This calculation gives correction to $\mu_{s}(q, 14)$
of the order of $2k_{B}T$, which results in $\mu_{s}(q, 14) 
\simeq - 6k_{B}T$. 

In Eq. (\ref{chempot}) we clearly see 
the origin of the $q$-dependence of $\mu_{s}(q, 14)$:
 proteins with larger $q$ stronger attract PE.
This dependence is of course translated  
in the $q$-dependence of $\mu_{s}(q,$ pH) and $N_1$.
Thus, all details of the function pI$(q, N)$ are
strongly $q$-dependent what 
makes possible to separate proteins with the same pI
and different hidden charges $q$.

Let us now estimate $\mu_{s}(q,$ pI).
First, we show that any strongly charged PE 
is actually quite rigid due to the Coulomb repulsion of its charges.
Consider the PE electrostatic 
tension force $F$, which makes a PE straight.
It depends on the linear density of charge $\eta$, screening radius $r_s$ 
and the PE persistence length $a$. The energy of PE of the 
length $L_{PE}$ is 
$L_{PE}(\eta^{2}/\varepsilon)\ln(r_s/a)$. Taking 
derivative with respect of $L_{PE}$
we get
\begin{equation}
F \simeq \frac{\eta^{2}}{\varepsilon}\ln\frac{r_s}{a}.
\label{force}
\end{equation}
Now we should recall that the density of protein surface charge 
fluctuates creating a bending force for a adsorbed PE. 
We assume that we are dealing with 
random charges with the density of charge of
each sign $q/4\pi R^2$ or the average distance 
between them $A=R(4\pi/q)^{1/2}$.

In order to find the energy of adsorption 
of a PE molecule in the PI point we should consider 
competition between the gain of the Coulomb energy 
which a PE molecule enjoys when it bends to reach 
negative charges and the loss of the elastic 
energy necessary to do that.
An important role is played here by the so called 
Larkin length\cite{Larkin}, $L$, which 
tells how long is the section of PE, which moves aside by 
the distance $r_s$
to use a typical fluctuation minimum of the 
electrostatic potential averaged over the stripe 
of length $L$ and the width $r_s$. This fluctuation 
of the potential equals
$-(e/\varepsilon r_s)(Lr_s/A^{2})^{1/2}(r_s/L)$.
Then a section of PE with the 
length $L$ gets the Coulomb energy 
$-(e\eta/\varepsilon) (Lr_s/A^{2})^{1/2}$ and 
the elastic energy $Fr_s^{2}/L$.
Thus the change of the total energy of PE equals
\begin{equation}
E \simeq \frac{L_{PE}}{L}
\left(-\frac{e\eta}{\varepsilon}
\frac{(Lr_s)^{1/2}}{A} 
+ \frac{Fr_s^{2}}{L}
\right).
\label{energy}
\end{equation}
Optimizing the right hand side of 
Eq. (\ref{energy}) with respect of $L$ 
and using Eq. (\ref{force})
we get $L \simeq r_s(A\eta/e)^{2/3}\ln^{2/3}(r_s/a)$.
For $q = 20A$, $A \simeq 0.8 R$ and 
for $\eta = e/l_B $ we get large $L \sim R$.
This results in quite small $E \simeq k_BT$.
The reason that $E$ is small
is the relatively small number of 
charges of the typical protein $q$ 
(and large distance between them $A$).

We emphasize, however, that $E$ 
is the attraction energy of a PE molecule to a typical spot
on the sphere. Actually, a PE molecule can find at the protein 
surface a spot, with anomalously large negative 
surface density of charge, where attraction is stronger.
Mentally moving a PE molecule along a protein globule surface
we can study distribution function of binding
energies, which has a lower energy tail and terminates
at the lowest energy which we want to find. 
The shape of the tail
(gaussian or exponential) and the lowest 
energy are strongly model dependent (see 
similar problems in semiconductor physics\cite{SE}).
The absolute minimum of energy can be larger than 
the energy $E$ in a typical spot
by the logarithm of the number of different independent states of a
PE on the surface of
the protein or by the square root of this logarithm.
This factor can reach 3 or 4 so that $|\mu_{s}(q, $pI$)|$
can be comparable to $|\mu_{s}(q, 14)|$, but
most likely it is smaller and the scenario B is realized.

For this scenario 
the concentration $N_2$ can be as small 
as 10 mM and the concentration $N_1$ can be 
10-50 times larger. It is clear
 from our estimates that 
both $N_1$ and $N_2$
depend on $q$ and this opens multiple 
possibilities of separation of proteins
with the same PI and different $q$.
 
We are grateful to A. Yu. Grosberg and T. T. Nguyen, 
for useful discussions. B. I. Shklovskii was supported
by NSF grant DMR-9985785. E. M. Baskin thanks the Center 
for Absorption in Science, Israel Ministry 
of Absorption for financial support.

\end{multicols}
\end{document}